\documentclass{optica-article}

\journal{opticajournal} % for journals or Optica Open

\articletype{Research Article}

\usepackage{amsmath,amssymb,verbatim,latexsym,mathrsfs,mathtools,amsthm,bbm,bm,hyperref,url,cancel,subcaption,enumitem,setspace}

\newcommand*{\bra}[1]{\langle #1\rvert}
\newcommand*{\ket}[1]{\lvert #1 \rangle}
\newcommand*{\braket}[2]{\langle #1 \lvert #2 \rangle}

\begin{document}

\title{Sub-Rayleigh Imaging of Unequal-Intensity Sources: Near-Quantum-Limit Multiparameter Estimation}

\author{Y. Batuhan Yilmaz,\authormark{1} Kent A. G. Bonsma-Fisher,\authormark{2}  Muhammad Mohid,\authormark{1} Noah Lupu-Gladstein,\authormark{3,4} Arthur O. T. Pang,\authormark{1} and Aephraim M. Steinberg\authormark{1,*}}

\address{\authormark{1}CQIQC and Department of Physics, University of Toronto, 60 Saint George St., Toronto, ON M5S 1A7, Canada\\
\authormark{2}Photonic Inc., Coquitlam, BC, Canada\\
\authormark{3}National Research Council of Canada, 100 Sussex Drive, Ottawa, Ontario K1N 5A2, Canada\\
\authormark{4}Department of Physics, University of Ottawa, 25 Templeton Street, Ottawa, Ontario, K1N 6N5 Canada}

\email{\authormark{*}steinberg@physics.utoronto.ca} %% email address is required; see note below about the corresponding author designation

% use {asbstract*} to suppress the copyright line. Copyright information will be added in production

\begin{abstract*} 
In optical imaging, diffraction strongly degrades the performance of conventional intensity-based estimation once the separation between two sources is below the Rayleigh-Abbe limit. Recent developments showed that this limitation 
can be surpassed using spatial-mode demultiplexing (SPADE), and this was demonstrated in several experiments for two equal-intensity spots of unknown separation. When there are multiple unknown parameters, as in the case of several
unequal sources with unknown intensities and unknown separation, cross-talk among the parameters makes the multi-parameter estimation problem significantly more challenging. In this paper, we adapt super-resolved position localization by inversion of coherence along an edge (SPLICE) to estimate both the separation and relative intensity of two incoherent sources simultaneously.  
We demonstrate a clear advantage over direct imaging (DI), achieving a root mean squared error (RMSE) approximately $50 \%$ larger than the quantum limit and a sixfold improvement over DI within the range of parameters we tested. This improvement can be even greater for smaller separations and larger intensity imbalances when crosstalk is suppressed.

\end{abstract*}

%%%%%%%%%%%%%%%%%%%%%%%%%%  body  %%%%%%%%%%%%%%%%%%%%%%%%%%
\section{Introduction}
For applications ranging from atomic physics and biology to satellite mapping and astronomy, researchers long believed that Rayleigh's criterion imposed a fundamental limit on the resolution of imaging systems \cite{rayleigh1879xxxi,abbe1873beitrage,abbe1882relation}. 
According to the Rayleigh's criterion, resolving two identical incoherent light sources requires their separation to be greater than the full-width half-maxima (FWHM) of their point spread functions (PSF). In our terms, ``resolving" two sources means estimating the separation between the two sources with a reasonably low uncertainty. 

 Conventional imaging methods, known as direct imaging (DI), estimate the intensity profile of the input light source. For two identical incoherent sources separated by $\delta$, the uncertainty in the estimated separation $\hat{\delta}$, measured using a diffraction-limited imaging system with a Gaussian point spread function (PSF) of width $\sigma$ and intensity-based detectors, can be approximated as $2\sqrt{2}\sigma^2/(\delta\sqrt{N})$ for $N$ detection events when $\delta \lesssim \sigma$ \cite{tsang2016quantum,ferretti2022quantum}. The resolution of DI is fundamentally limited by diffraction from a finite aperture. When the full width at half maximum (FWHM) of the PSFs exceeds the source separation, the combined intensity profile becomes difficult to distinguish from that of a single source. Although larger apertures can improve the resolution, constructing larger telescopes may be cost-prohibitive \cite{schmidt1997telescope} or infeasible in certain scenarios, particularly microscopy.  

Breakthroughs in surpassing DI have been achieved using super-resolution techniques that modify the illumination source or utilize the nonlinear response of the object being imaged \cite{hell1994breaking,betzig2006imaging,hess2006ultra,hemmer2012universal,giovannetti2004quantum,Tamburini2006vortice,Tsang2009centroid,rozema2014scalable,Giovannetti2009sub,shin2011quantum,schwartz2013superresolution}.  However, such approaches are not applicable in scenarios where the illumination and the object response cannot be controlled, with astronomy being a prominent example. More recently, theory \cite{tsang2016quantum,lupo2016ultimate,rehacek2017optimal,gefen2019overcoming} and experiments \cite{paur2016achieving,yang2016far,tang2016fault,tham2017beating,donohue2018quantum,zhou2019quantum,howell2023super,frank2023passive,rouviere2024ultra} have shown that the separation of two incoherent, equal-intensity sources can be estimated beyond the resolution achievable by DI using only passive linear-optical devices. This is possible because light, as a complex field with both amplitude and phase, encodes more spatial information than is accessible through direct imaging (DI), which measures only the intensity distribution and is inherently insensitive to phase.  Phase-sensitive schemes, such as spatial-mode demultiplexing (SPADE) \cite{tsang2017subdiffraction}, recover this lost information by projecting the light onto Hermite-Gaussian (HG) modes and achieve constant uncertainty $2\sigma/\sqrt{N}$ \cite{tsang2016quantum}. This result implies that, in principle, arbitrarily small separations can be resolved with the same precision as large ones.

HG mode sorters are not necessary to achieve sub-Rayleigh resolutions. The super-resolving position localization by inversion of coherence about an edge (SPLICE) \cite{tham2017beating} method uses a phaseshifter that implements a $\pi$ phase shift to half of the transverse profile of the incoming light, followed by a single mode fiber centered at the region of phase discontinuity of the phaseshifter. This configuration approximates a projector on the first Hermite-Gauss mode $\text{HG}_1$. Providing an experimentally simpler alternative, SPLICE can achieve nearly 2/3 of the theoretically allowed maximum information.

All of the work mentioned so far concerns single-parameter estimation. However, many real-world systems involve multiple unknown parameters. For instance, a star with exoplanets  involves parameters such as the planets’ relative intensities and their distances from the star.
 Even in the simple case of a single exoplanet, detection and parameter estimation are challenging due to the extreme brightness imbalance between the star and the planet \cite{traub2010direct,huang2021quantum,deshler2024achieving,radhakrishnan2025towards}. When the intensity ratio is unknown, the uncertainty of the estimator $\hat\delta$ estimated with DI scales as 
$1/\delta^2$, which is worse than the 
$1/\delta$ scaling of the equal-intensity case. Moreover, the parameters cannot be measured independently and must be estimated jointly. These difficulties make systems with multiple unknown parameters particularly challenging to characterize. Theoretical studies on multi-parameter estimation \cite{Soto2017multiparam,bonsma2019realistic,PhysRevA.100.032104,prasad2020quantum,xin2021optical,katamadze2023breaking} have shown that phase-sensitive measurements can achieve a higher precision than DI. There have been experiments that estimated either separation of relative intensities of two sources using multi-plane light converters (MPLC) \cite{santamaria2023spatial,santamaria2024single}. 
Multi-parameter estimation has been demonstrated in the temporal domain where the temporal separation, relative intensity and centroid of two ultra-short pulses were estimated\cite{ansari2021achieving}. Recently, an experiment using multi-plate light converters (MPLCs) demonstrated the simultaneous estimation of the separation and relative intensity of two sources with a known center of mass \cite{wallis2025spatial}. A similar experiment demonstrated the simultaneous estimation of the separation, relative intensity and the centroid of two sources with an unknown center of mass 
\cite{grateau2026multiparameter}. 

 In this paper, we extend the SPLICE method \cite{bonsma2019realistic} to simultaneously estimate the relative intensity and separation along a known axis between two sources with a Gaussian PSF. Because the optimal measurement basis for a Gaussian PSF is the interferometric Hermite-Gauss ($\textnormal{iHG}$) basis \cite{rehacek2017optimal}, we apply two projectors approximating $(\textnormal{HG}_1(x)\pm \textnormal{HG}_2(x))/\sqrt{2}$ to approach the quantum limit in accuracy as closely as possible. By implementing SPLICE in the single-photon regime, we rigorously evaluate our scheme against shot-noise-limited bounds, demonstrating performance significantly superior to $\textnormal{DI}$ and reaching within $\approx 50\%$ of the quantum limit for both separation and relative intensity. Finally, we analyze how this advantage scales with source separation, intensity balance, and apparatus extinction ratio (cross-talk).

The rest of this paper is organized as follows: Section \ref{sec:theory} establishes the information-theoretic precision bounds for our experiment, Section \ref{sec:exp} explains how the experiment is conducted, Section \ref{sec:res} presents and discusses the main results, and Section \ref{sec:conc} provides concluding remarks.

\section{Theory}\label{sec:theory}
\begin{figure}
    \centering
    \includegraphics[width=0.9
 \textwidth]{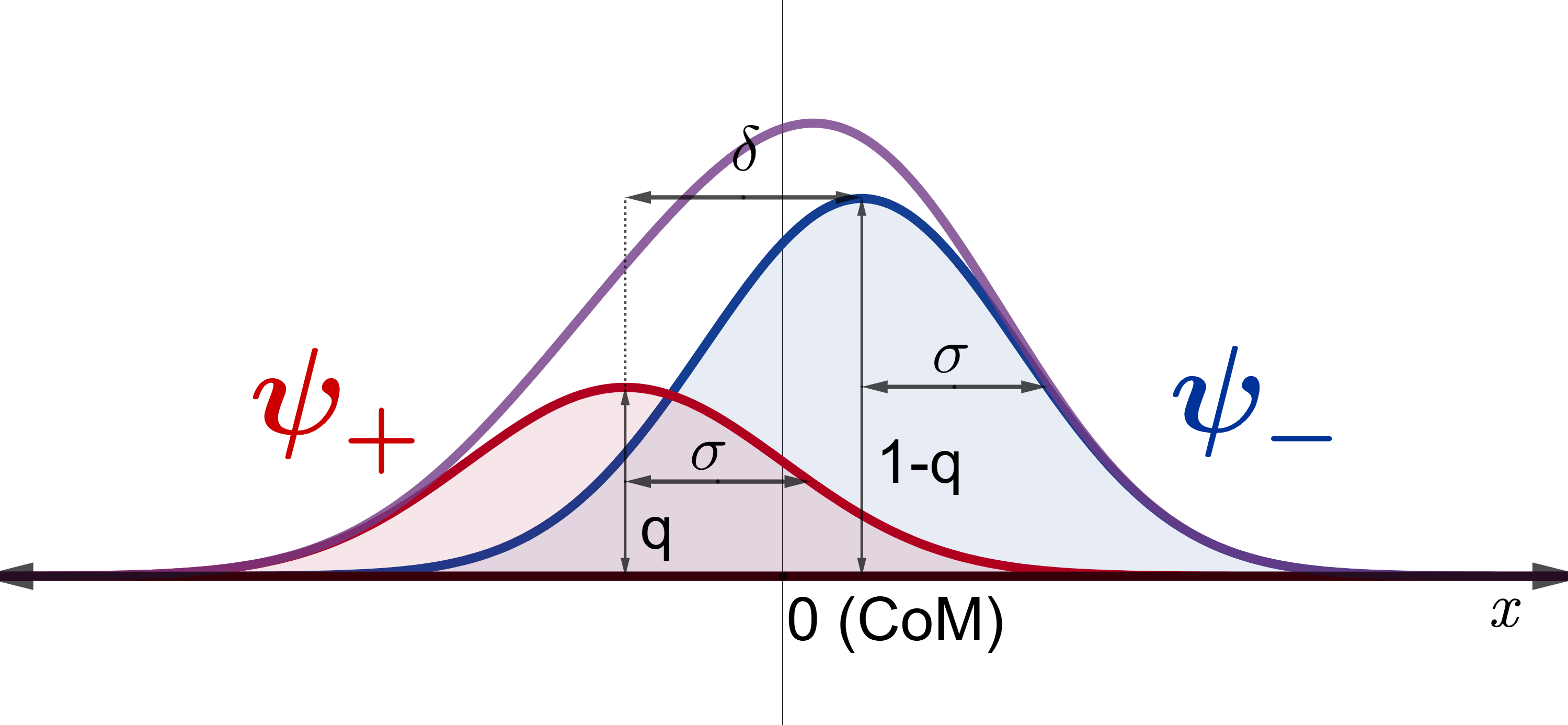}
    \caption{Illustration of the PSF of the two sources with separation $\delta$ and relative intensity $q$ in the image plane (red and blue) and the PSF of the two sources combined (purple). The axis is centered on the combined PSF's center of mass (CoM). }
    \label{fig:inputsketch}
\end{figure}

 We consider two incoherent sources with relative intensities $q$ and $1-q$ separated by a distance $\delta$ about a midpoint $c$. The midpoint $c$ can be expressed in terms of the center of mass $\mu$ (first moment of the intensity distribution), $\delta$, and $q$:
 \begin{equation}
     c= \mu +(1-2q)\delta/2.
 \end{equation}
 The center of mass can be easily and efficiently estimated using standard DI methods \cite{tsang2017subdiffraction}. Therefore, the primary challenge lies in estimating the separation $\delta$ and the relative intensity $q$, which constitute the main focus of our experiment. Throughout this work, we assume that the center of mass $\mu$ is known.

If the probability of detecting a photon within a coherence time is very small, almost all of the detection events consist of single photon events \cite{tsang2016quantum}. In this case, the electromagnetic field can be approximated as a statistical mixture of single-photon states emitted from either source,
\begin{equation}\label{rho_def}
    \rho \approx q \ket{\psi_+}\bra{\psi_+}+(1-q) \ket{\psi_-}\bra{\psi_-},
\end{equation}
where $\ket{\psi_\pm}$ denotes the wavefunction of a photon emitted from one source or the other. The wavefunction $\ket{\psi_\pm}$ can be decomposed in the spatial basis $\{\ket{x}\}_{x\in \mathbb{R}  }$:

\begin{equation}
  \ket{\psi_\pm}= \int dx  \ket{x} \braket{x }{ \psi_\pm}   = \int dx \psi_\pm(x) \ket{x} 
\end{equation}

The image of an ideal point source going through the apparatus has a PSF with a finite width depending on the properties of the apparatus. Throughout this work, we assume a Gaussian PSF, 
\begin{equation}\label{gauss}
    \ket{\psi_\pm}=\int dx \frac{1}{(2 \pi \sigma^2)^{1/4}} e^{-\frac{(x-(c\pm\delta/2))^2}{4\sigma^2}} \ket{x}.
\end{equation}
 Many realistic PSFs can be reasonably approximated as Gaussian. This approximation also simplifies the analytical calculations.

Measuring the state with an n-outcome POVM $\{M_k\}_{k \in {1,...,n}}$ yields measurement outcomes $k \in {1,...,n}$ with the corresponding probabilities,
\begin{equation}
    P_k= \text{Tr}( M_k \rho).
\end{equation}
One can infer the values of the parameters $\delta$ and $q$ from the measurement outcomes with a minimum achievable uncertainty set by the Cram\'er-Rao bound \cite{rao1945information}:
\begin{equation}
    \Sigma \geq F^{-1},
\end{equation}
where, $\Sigma$ is defined as the covariance matrix for the estimators $\hat\delta$ and $\hat q$, $\Sigma_{ij}=\text{Cov}(i,j)$, $i,j \in \{\hat\delta,\hat q\}$  and $F$ is the Fisher Information matrix \cite{ly2017tutorial} defined as
\begin{equation}\label{Fij}
     F_{ij}=\sum_m \frac{\partial_{i} P_m \partial_{j} P_m}{P_m},
 \end{equation}
where $i,j \in \{\delta, q\}$. The inequality $\Sigma \geq F^{-1}$ for the two matrices means that the matrix $\Sigma -F^{-1}$ is positive semi-definite. 

 The Cram\'er-Rao bound depends on the choice of POVMs. In many cases, one is interested in the quantum Cram\'er-Rao bound (QCRB), which sets a bound for the best uncertainty given the best choice of POVM:
\begin{equation}
    \Sigma \geq F^{-1} \geq Q^{-1}.
\end{equation}
Q is the quantum Fisher Information matrix defined as \cite{helstrom1969quantum,liu2020quantum},
\begin{equation}\label{eqn:qfidef}
    Q_{ij}=\frac{1}{2}Tr(\rho(L_iL_j+L_jL_i)).
\end{equation}
Here $L_i$ is the symmetric logarithmic derivative operator \cite{helstrom1967minimum}, implicitly defined by
\begin{equation}
    \partial_i \rho=\frac{1}{2}( L_i\rho+\rho L_i).
\end{equation}
\subsection{Quantum Limit}
 Calculating the QCRB for the estimators $\hat \delta$ and $\hat q$ using the expressions in \cite{Soto2017multiparam} for a Gaussian PSF yields
\begin{equation}\label{eqn:qdd}
    \text{Var}[\hat \delta] > (Q^{-1})_{\delta \delta}= \frac{\sigma^2}{N}\frac{1-e^{-\Delta^2}-q(1-q)4\Delta^2 e^{-\Delta^2}}{q(1-q)(1-e^{-\Delta^2}-\Delta^2 e^{-\Delta^2})},
\end{equation}
\begin{equation}\label{eqn:qqq}
    \text{Var}[\hat q] > (Q^{-1})_{qq}= \frac{q(1-q)(1-\Delta^2 e^{-\Delta^2})}{N(1-e^{-\Delta^2}-\Delta^2 e^{-\Delta^2})},
\end{equation}
where $\Delta = \delta/(2\sigma)$, $N$ is the number of detections, and $Q^{-1}$ is the inverse of the quantum Fisher
information matrix defined in Eq. \ref{eqn:qfidef}. The bound for separation $\delta$ reduces to $(Q^{-1})_{\delta \delta}=4 \sigma^2$ for balanced sources, as expected \cite{tsang2016quantum}. For unbalanced sources $(q\neq 0.5)$ and small separations ($\delta \ll \sigma$), we can simplify the expressions in Eq. \ref{eqn:qdd} and Eq. \ref{eqn:qqq}, to
\begin{equation}\label{CRBqdd}
        (Q^{-1})_{\delta \delta}\approx \frac{8 (1-2q)^2\sigma^4}{N q(1-q) \delta^2},
\end{equation}
and
\begin{equation}\label{CRBqqq}
    (Q^{-1})_{qq}\approx \frac{32q(1-q)\sigma^4}{N\delta^4},
\end{equation}
respectively. Both variances diverge when $\delta$ approaches zero. In reality, the variances of our estimators cannot be infinite because the parameters we estimate are bounded. This discrepancy occurs because CRB relies on the assumption of unbiasedness of the estimators, which is true in the asymptotic limit, when the number of measured photons is sufficiently large. The relative intensity $q$ has a value between 0 and 1. Therefore, its uncertainty cannot be greater than 1. Similarly, the estimator for separation $\delta$ converges to a finite value in the low photon-number regime \cite{bonsma2019realistic}. Therefore, we used Monte Carlo simulations to compare our experimental data in the biased regime.

 The projectors implemented by the measurements must be linearly independent and orthogonal to $\textnormal{HG}_0(x)$, otherwise, the contribution from $\textnormal{HG}_0(x)$ dominates the measurement result instead of the higher-order terms. The optimal strategy to estimate $\delta$ and $q$ 
 in our scenario is to implement interferometric Hermite-Gauss $(i\textnormal{HG})$ measurements  \cite{tsang2017subdiffraction} by projecting on $(\textnormal{HG}_1(x)\pm \textnormal{HG}_2(x))/\sqrt{2}$ with the center of mass of the sources set as the origin. The probabilities of photon detection with $i\textnormal{HG}$ projectors are:
\begin{equation}
P_{i\textnormal{HG}\pm}=\frac{q(1-q)\delta^2}{8\sigma^2}\pm\frac{ q(1-q)(1-2q)\delta^3}{8\sigma^3}+O(\delta^4)
\end{equation}
\subsection{SPLICE}
In principle, the quantum limit can be achieved with an ideal mode sorter. Remarkably, even simpler optical setups can approximate the \textnormal{HG} projectors and achieve superresolution \cite{bonsma2019realistic}. Our measurement setup consists of a fiber collimator which works as a projector in $\textnormal{HG}_{0}(x)$ and a pair of microscope cover slips (MCS) one of which is tilted to apply relative $\pi$ phaseshift. Therefore, the projectors we can apply are in the following form where $f$ is the position of the fiber collimator and $g$ is the position of the phaseshifter:
\begin{equation}
    \ket{\phi}=\int_{-\infty}^{\infty} dx  \phi(x) \ket{x}
\end{equation}
\begin{equation}\label{proj}
    \phi(x)=\braket{x}{\phi}=\frac{sgn(x-g)}{(2 \pi \sigma^2)^{1/4}} e^{-\frac{(x-f)^2}{4\sigma^2}}
\end{equation}
While in reality, the expression for the projectors is more complicated due to imperfections in the fiber collimator and the MCSs, this expression is a reasonably good model to analyze the performance of our scheme. 
The Fisher Information matrix for a set of M two-outcome projectors $\{\phi_m(x)\}_{m \in 1,...,M}$ yields:
\begin{equation}\label{fproj}
    J_{ij}= \frac{1}{M}\sum_{m=1}^M \frac{ \partial_iP_m(\delta,q,c)\partial_jP_m(\delta,q,c)}{P_m(\delta,q,c)(1-P_m(\delta,q,c))}
\end{equation}
Where $P_m(\delta,q,c)$ is defined as:
\begin{equation}\label{Pproj}
P_m(\delta,q,c)=q\braket{\psi_+}{\phi_m}\braket{\phi_m}{\psi_+}+(1-q)\braket{\psi_-}{\phi_m}\braket{\phi_m}{\psi_-}
\end{equation}
  The optimal pair of projectors $\phi_\pm(x)$ that minimize the uncertainty of $\delta$ and $q$ is implemented by setting the MCS position $g=\pm\sigma$ and the fiber collimator position $f=\pm2\sigma$, where the center of mass is defined as the origin ($\mu=0$). Although these two projectors can be thought of as approximations of the modes $\frac{1}{\sqrt{2}} (\textnormal{HG}_{1}(x)\pm \textnormal{HG}_{2}(x))$, they are not optimized to maximize similarity, they are optimized to minimize uncertainty \cite{tham2020quantum}. The probabilities of photon detection are: 
\begin{figure}
    \centering
    \includegraphics[width=0.9
 \textwidth]{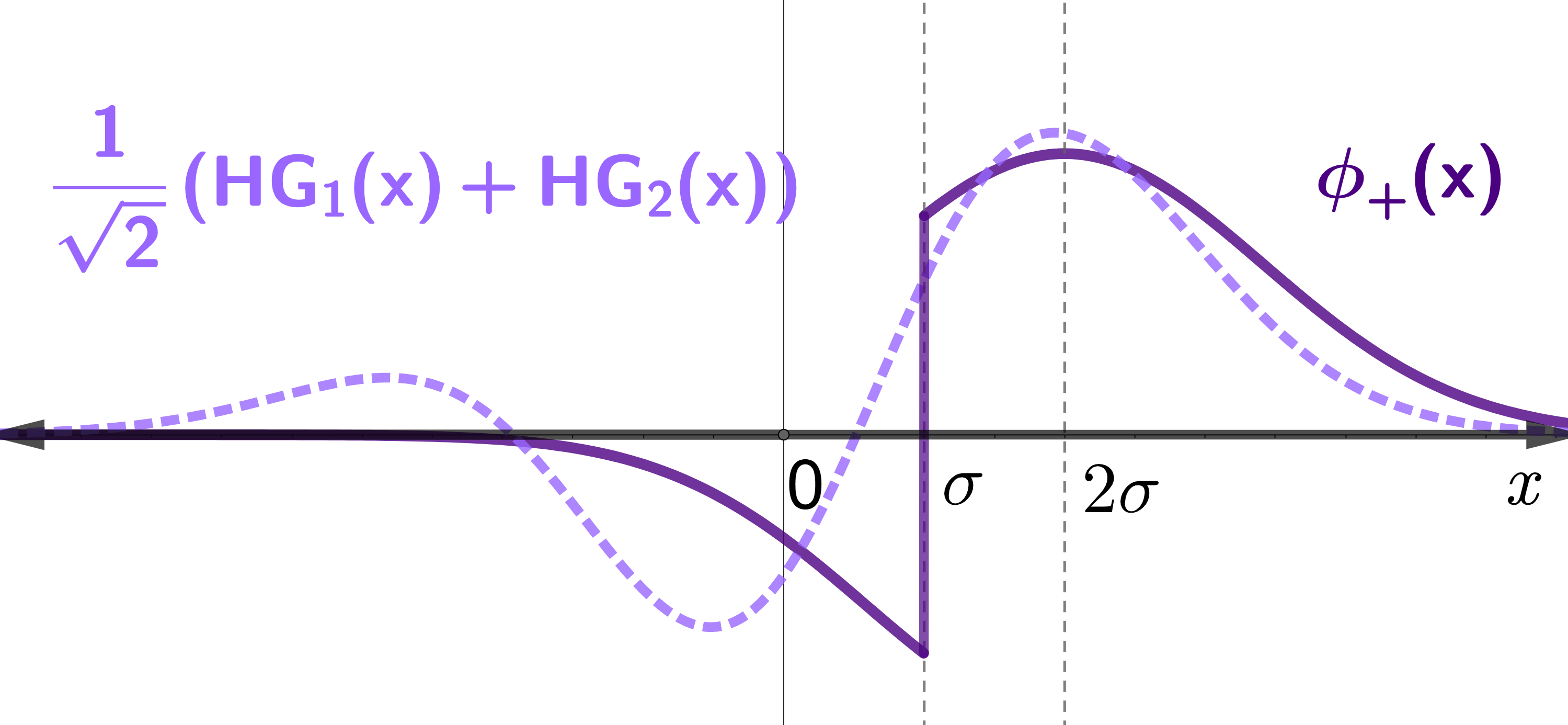}
    \caption{Illustrations and the comparison of the ideal $(\textnormal{HG}_1(x)+\textnormal{HG}_2(x))/\sqrt{2}$ projector (dashed line) and the projector implemented in the experiment, $\phi_+(x)$ (solid line). The center of mass of the two sources is defined as the origin. The projector $\phi_+(x)$ is applied placing the phase shifter to $x=\sigma$ and placing the fiber collimator to $x=2\sigma$. }
    \label{fig:HG12sketch}
\end{figure}

\begin{equation*}\label{Pprojpm}
    P_{\phi_\pm} = \frac{q(1-q)\delta^2}{2 \pi e \sigma^2}\pm \frac{q(1-q)(1-2q) \delta^3}{2 \pi e \sigma^3} + O(\delta^4).
\end{equation*}
For small  $\delta$ and $q$, $P_{\phi_\pm}$  is approximately proportional to $P_{i\textnormal{HG}_\pm}$, differing only by a scaling factor of $4/(\pi e)\approx 0.46$. The CRB for the projectors $P_{\phi_\pm}$ is the following:
\begin{equation}\label{CRBspldd}
    \Sigma^{SPL}_{\delta \delta} > (J^{-1})_{\delta \delta}=\frac{ 2\pi e(1-2q)^2\sigma^4}{N  q(1-q) \delta^2}
\end{equation}
\begin{equation}\label{CRBsplqq}
    \Sigma^{SPL}_{qq} > (J^{-1})_{qq}=\frac{8 \pi e q(1-q)\sigma^4}{N\delta^4}
\end{equation}
 This extra factor of $4/(\pi e)$ carries over to the CRB for SPLICE, meaning that with SPLICE, we can extract $46\%$ of the total information available according to the quantum limit. For equal intensities, the CRB for $\delta$ becomes constant $(J^{-1})_{\delta \delta}=2\pi e \sigma^2 $.  The ratio of CRBs for equal intensities $(Q^{-1}(q=0.5))_{\delta \delta}/(J^{-1}(q=0.5))_{\delta \delta}=2/(\pi e)\approx 0.23 $ is worse because we optimized SPLICE to perform the best at unequal intensities. 
\subsection{Direct Imaging (DI)}
 DI extracts the properties of the light source from its intensity profile:  \begin{equation}\label{DI} P_{DI}(x)=\bra{x}\rho\ket{x}.
 \end{equation}
 In reality, the pixel size of the camera will limit the resolution of the measurements. However, we can neglect that and consider a continuous probability distribution to calculate the absolute limit for the precision. The bounds that are derived with this assumption can be achieved with cameras with pixel widths much smaller than the separation between the sources. The FI matrix with this assumption can be calculated as:

 \begin{equation}\label{Fij_cont}
     F_{ij}=\int dx \frac{\partial_{i} P_{DI}(x) \partial_{j} P_{DI}(x)}{P_{DI}(x)}
 \end{equation}
For unbiased estimators, the covariance matrix for the parameters is bounded by the Cram\'er-Rao bound (CRB):
\begin{equation}\label{CRB}
    \Sigma^{DI} \geq F^{-1}
\end{equation}
For small separations $(\delta \ll \sigma)$ and unequal sources $(q \neq 0.5)$ variances for $\delta$ and $q$ that can be achieved by DI are approximately bounded by:
\begin{equation}\label{CRBdidd}
    \Sigma^{DI}_{\delta \delta} > (F^{-1})_{\delta \delta} = \frac{6 (1-2q)^2\sigma^6}{N q^2(1-q)^2 \delta^4}
\end{equation}
\begin{equation}\label{CRBdiqq}
    \Sigma^{DI}_{qq} > (F^{-1})_{qq} = \frac{24\sigma^6}{N\delta^6}
\end{equation}
The bound given above for $\delta$ is valid for $q\neq 0.5$, unbalanced sources. For equal sources the scaling changes to $\delta^{-2}$: 

\begin{equation}\label{CRBdidd_eq}
    (F(q=0.5)^{-1})_{\delta \delta} =  \frac{8\sigma^4}{N \delta^2}
\end{equation}

\begin{figure}
    \centering
    \includegraphics[width=0.9
 \textwidth]{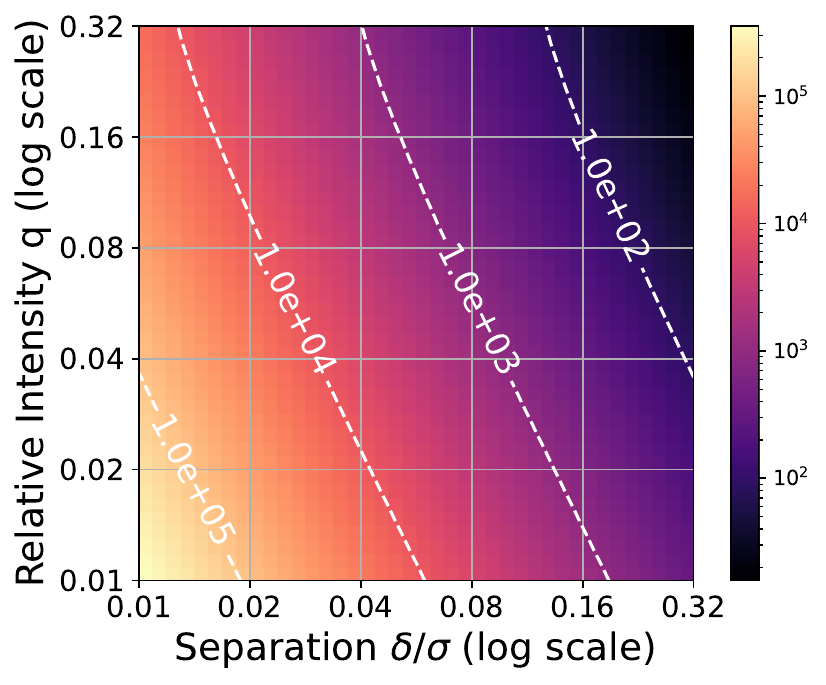}
    \caption{Ratio of the CRBs for $\delta$ with DI and SPLICE  in a log-log contour plot. The advantage of SPLICE grows as $\delta$ and $q$ get smaller.}
    \label{fig:ratiocrb}
\end{figure}

 To compare DI to SPLICE, we can calculate the ratio of the bounds of the two methods, as calculated below and plotted in Fig. \ref{fig:ratiocrb}.

\begin{equation}\label{eqn:ratiospl}
    R= \frac{(F^{-1})_{\delta \delta}}{(J^{-1})_{\delta \delta}}= \frac{(F^{-1})_{qq}}{(J^{-1})_{qq}} = \frac{3 \sigma^2}{\pi e q(1-q) \delta^2}
\end{equation}

SPLICE offers a significant advantage over DI for both estimating $\delta$ and $q$.  There is a quadratic improvement in the scaling of the variance of $\delta$ with respect to $\delta$ and $q$ for small values of $\delta$. The variance of $q$ diverges with a lower power of $\delta$ with SPLICE and decreases proportionally with $q$. 

\section{Experiment} \label{sec:exp} 

\begin{figure}
    \centering
    \includegraphics{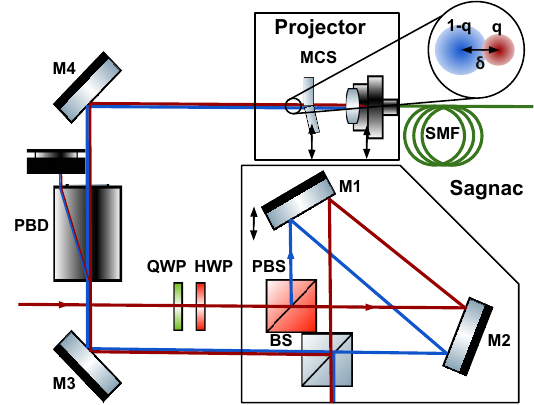}
    \caption{The experimental setup. The spatial profile of the two sources is illustrated on the top right. The Sagnac configuration prepares the two beams with unequal intensities and the fiber collimator with microscope cover slips (MCS) applies the projectors $\phi_\pm(x)$. We use the polarization degree of freedom of light to control the relative intensity. The quarter-wave plate (QWP) and half-wave plate (HWP) sets the polarization. The polarizing beamsplitter (PBS) at the beginning of the interferometer splits the horizontal and the vertical portions of the light, and the non-polarizing beamsplitter recombines the two beams. The distance between two beams is controlled by the mirror M1 mounted on a translation stage. To eliminate the polarization difference, the beams go through a polarizing beam displacer (PBD) set to 45 degrees. The fiber collimator and the pair of microscope cover slips (MCS) on translation stages realize the projector by applying a $\pi$ phaseshift on part of the transverse plane.  }
    \label{fig:setup}
\end{figure} 
 We used the setup shown in Fig \ref{fig:setup} for the experiment. Our apparatus mimics the broadened PSF of two point sources in the image plane using two laser beams with a beam radius $\sigma = 397 $ µm and then applies the projectors $\ket{\phi_\pm}$. We used an $810$ nm Ti:sapphire laser with a pulse repetition rate of $82$ MHz, and a pulse duration of $140$ fs. First, the light enters the Sagnac configuration which prepares two beams with desired separation and relative intensity. The ratio of horizontally and vertically polarized light, controlled by the HWP, determines the splitting ratio, i.e. the relative intensity $q$. The position of mirror M1 along the translation-stage axis determines the beam separation $\delta$. The path length difference in the interferometer is much longer than the coherence length of the laser ($42$ µm), ensuring that the beams are incoherent. After the beams exit the interferometer, they pass through a polarizer set to 45 degrees to eliminate any possibility of polarization-related distinguishability. Then, the beams go through the measurement stage, which consists of a pair of No. 1 ($130-160$ µm) microscope cover slips (MCS) and a fiber collimator, both of which are mounted on translation stages. One of the MCSs is tilted to create a relative $\pi$ phaseshift in the transverse plane. After the fiber collimator, the light goes through a pair of free-space ND filters and is attenuated to 1,600,000 photons per second on average. The attenuation stage is added to evaluate the performance of our method in the shot-noise-dominated regime and to make a clear comparison to the theoretical limits. The attenuated beams were measured by a fiber-coupled SPCM with 250-300 dark counts per second.
We used the following calibration procedure. We move the MCSs such that the interface between the two slips is $10 \sigma$ away from the geometric center of the beams so that the beams go through the same slip with no relative phase shift. We move M1 to reduce the separation to zero. We measure the beam intensities one by one and find the waveplate angles that set the desired relative intensities. To calibrate the separation, we set M1 to a certain position, set $q=0$ and scan the left beam with the fiber collimator, and set q=1 and scan the right beam with the fiber collimator. These scans give us the beam positions and separations. The separation values are repeatable within an uncertainty of 5 µm, and the relative intensity values are repeatable to within $1\%$ uncertainty.

 We initially align our measurement stage to the center of mass of the beams, which is defined as the position $x=0$. We set the MCSs to $x=\pm\sigma$ and the fiber collimator to $x= \pm 2 \sigma$ to realize the projectors $\phi_\pm(x)$. We measure 10 different values of separation and 9 different values of relative intensity, corresponding to a total of 90 settings, with 20 trials performed at each setting. For each trial, measurements are collected over 2 seconds, with photon counts recorded every 0.1 seconds. 

The projectors we picked for our protocol are ideally orthogonal to the $\textnormal{HG}_{0}(x)$ mode. However, experimental imperfections cause a non-zero coupling from $\textnormal{HG}_{0}(x)$ and other unwanted modes. Since the dominant mode is $\textnormal{HG}_{0}(x)$ in the small separation regime, we can neglect the cross-talk from other higher-order modes. The way to model the cross-talk is to use an imperfect projector $\ket{\Psi_\epsilon}\bra{\Psi_\epsilon}=(1-\epsilon)\ket{\Psi_0}\bra{\Psi_0}+\epsilon\ket{\textnormal{HG}_0}\bra{\textnormal{HG}_0}$, where $\ket{\Psi_0} \bra{\Psi_0}$ would correspond to an ideal projector. The photon detection probability can be modified in the following way to include the effect of the cross-talk as a background term:

\begin{equation}\label{Pprojpm_eps}
    P_{\phi_\pm} \approx \epsilon+\frac{q(1-q)\delta^2}{2 \pi e \sigma^2}\pm \frac{q(1-q)(1-2q) \delta^3}{2 \pi e \sigma^3} + O(\delta^4)
\end{equation}

If the cross-talk is comparable to the original signal, the scaling advantage with respect to $\delta$ and $q$ is replaced by a constant inversely proportional to the cross-talk parameter, $R \approxeq 3/(2e\pi^2  \epsilon)$ in the unbiased regime.  

The cross-talk from $\textnormal{HG}_{0}(x)$ for both projectors is around $0.1-0.2\%$ in our experiment. The main sources of cross-talk are deviations of the beams from normal incidence ($\sim 20$  µrad), the air gap between the MCSs ($\sim 10$ µm) and limited accuracy of the translation stages ($0.7$  µm).  In addition, the two input beams have slightly different incidence angles ($\sim 5$  µrad), which is the best that can be achieved with our current capabilities. This difference in incidence angles leads to different cross-talk levels for each beam. This discrepancy can be incorporated in our model as the following:

\begin{equation}\label{Pprojpm_eps1}
    P_{\phi_\pm} \approx \epsilon_0+\epsilon_1 q+\frac{q(1-q)\delta^2}{2 \pi e \sigma^2}\pm \frac{q(1-q)(1-2q) \delta^3}{2 \pi e \sigma^3} + O(\delta^4)
\end{equation}

The effect of the $q$-dependent background is discussed in the next section.

\section{Results and Discussions} \label{sec:res}

\begin{figure*}
    \centering
    \includegraphics[width=\textwidth]{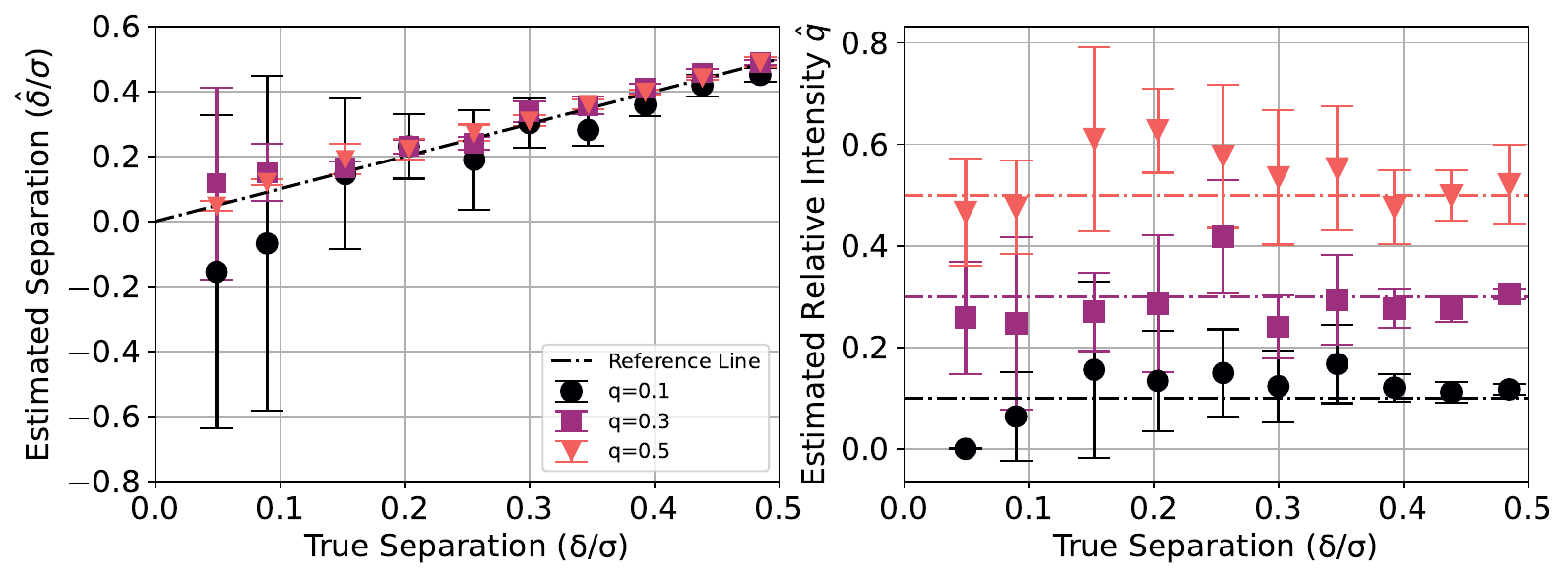}
    \caption{Estimated values vs. the true values of the parameters. Different colors are the results for different relative intensities. The separation values are normalized by the width of the Gaussian PSF $\sigma = 397$ µm. The reference lines show the true values of the parameters. Each point is the average of 10 trials. Each estimate uses on average $320{,}000$ single photon detections. The error bars represent the standard deviation of the 10 trials of each setting.}
    \label{fig:Ests}
\end{figure*}

\begin{figure*}
    \centering
    \includegraphics[width=\textwidth]{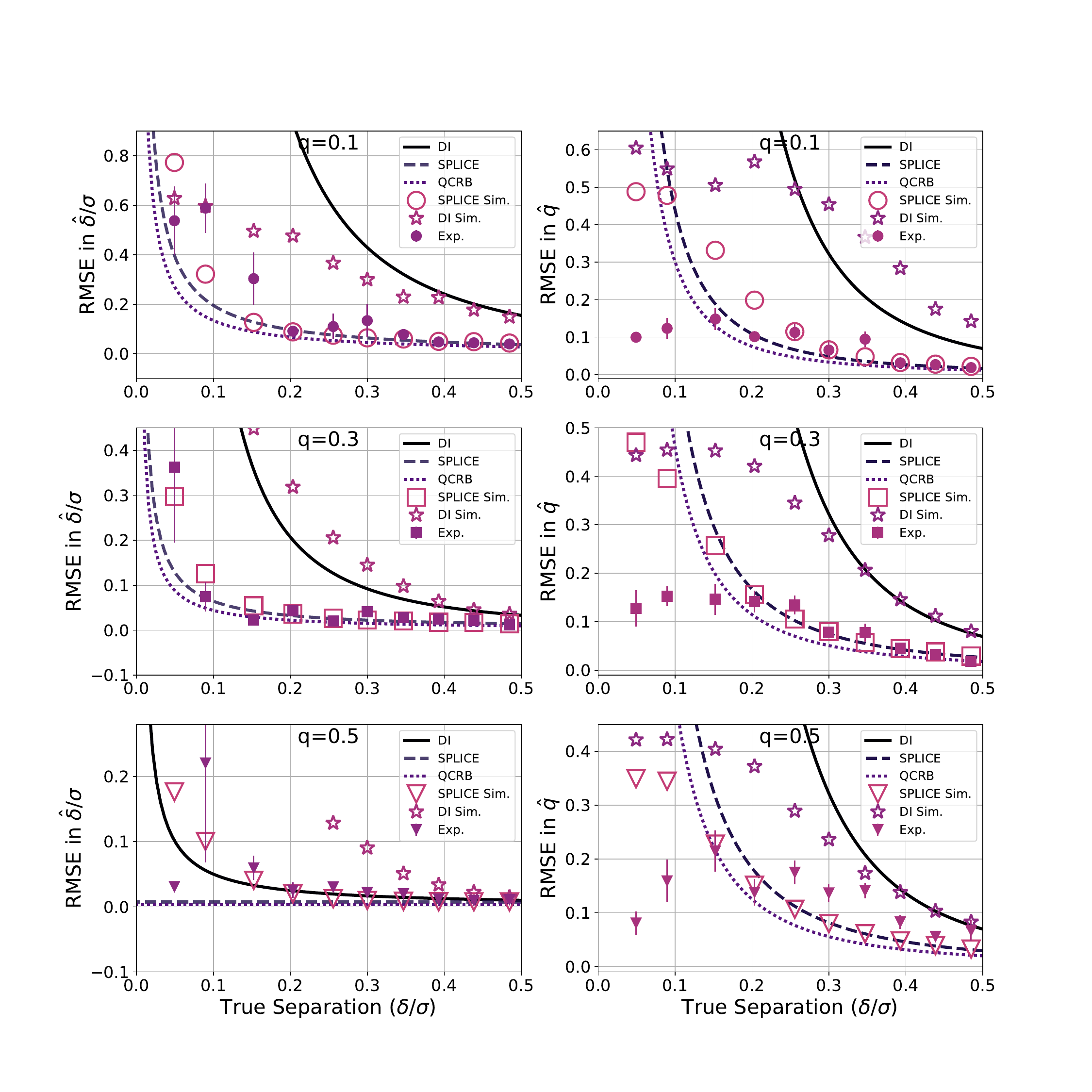}
    \caption{Average of the root mean squared error (RMSE) of the estimated parameters. The separation values are normalized by the width of the Gaussian PSF $\sigma = 397$ µm. Each point shows the average RMSE computed over 20 independent random partitions of the data into calibration and evaluation sets. The majority of the data points (solid markers) closely follow the theoretical limit for SPLICE (dashed lines) and perform better than the theoretical limit for direct imaging (DI) (solid lines). The hollow stars are the simulations for DI, and the other hollow markers are simulations for SPLICE. Each hollow marker is an average of 1000 simulated trials with an average of $320{,}000$ simulated single photon detections at each trial. }
    
    \label{fig:RMSE}
\end{figure*}

We used a constrained least-squares estimator to infer the separation $\delta$ and the relative intensity $q$ from the measurement results. Figure \ref{fig:Ests} shows estimated values of $\delta$ and $q$ as we scan the true  separation for 3 different true values of $q$. Figure \ref{fig:RMSE} shows the comparison of experimental root mean squared error (RMSE) to the theoretical limits and simulations. Each data point represents the average of 10 trials, each trial consists of $320{,}000$ single photon detections. The error bars in Figure \ref{fig:Ests} are the standard deviations of each of the 10 trials. The dashed lines in Figure \ref{fig:Ests} show the true values of the parameters. Estimated separations closely follow the reference line until $\delta/\sigma \leq 0.1$ where the uncertainty in estimated value becomes large for $q = 0.1$. The uncertainties can be reduced by increasing the measurement time and detecting more photons.

To estimate the parameters robustly against the experimental imperfections, we calibrated our measurement setup by fitting polynomials to its responses using the function below inspired from the physical model:

\begin{equation*}
    P_{f} =\epsilon_0 + \epsilon_1 q + c_2\frac{q(1-q)\delta^2}{2 \pi e \sigma^2} + c_3\frac{q(1-q)(1-2q) \delta^3}{2 \pi e \sigma^3} 
\end{equation*}

\begin{equation}
    + \frac{q(1-q)(c_{40}+c_{41}q+c_{42}q^2)\delta^4}{12 \pi e \sigma^4}
\end{equation}

Of the 20 trials at each setting, 10 are used to calibrate the free parameters of the model fitted to the experiment, while the remaining 10 are used to estimate the parameters $\delta$ and $q$. The free parameters $\epsilon_0$, $\epsilon_1$, $c_2$, $c_3$, $c_{40}$, $c_{41}$, and $c_{42}$ incorporate the imperfections of the setup. They are estimated with a maximum likelihood procedure for each of the projectors $\phi_+(x)$ and $\phi_-(x)$, with $\chi^2$ values of 1.46 and 1.93, respectively. We use half of the trials for this calibration, and we use the other half to test the RMSE of our protocol. The parameters $\epsilon_0$ and $\epsilon_1$ correspond to the cross-talk between $\textnormal{HG}_0(x)$ and the projectors $\phi_\pm(x)$. 

 To evaluate estimation performance, the data were repeatedly partitioned into calibration and evaluation sets. In each repetition, 10 trials were randomly selected to fit the model's free parameters, while the remaining 10 trials were used for parameter estimation. The RMSE was computed from the resulting estimates. This procedure was repeated 20 times with different random partitions, and the reported RMSE corresponds to the mean across repetitions, shown in Figure \ref{fig:RMSE}. Error bars indicate one standard deviation of the RMSE distribution. 
 
 The RMSEs in Figure \ref{fig:RMSE} mostly follow the CRB derived in Section \ref{sec:theory} and demonstrate an advantage that increases with decreasing $\delta$ and $q$. The CRBs becoming very large at small $\delta$ and $q$ mean that no meaningful information can be extracted and the actual value of CRB loses validity because the estimators become biased. In those regimes, it is more helpful to compare our results to simulations to understand how advantageous our method can be in practice. 

We included Monte Carlo simulations for DI for the same number of detection events to compare our results with the biased regime of DI.  Each hollow marker is an average of 1000 simulations that use the method of moments to estimate the parameters. In all of the plots, There are regions where SPLICE has a much smaller RMSE than DI can achieve. The observed advantage is at most around 5-6 times. The highest advantage possible with the crosstalk ($\epsilon=10^{-3}$) in our setup is $\sqrt{3/(2e\pi^2\epsilon)}=7.5$.   

%For $q=0.5$, the theoretical RMSEs become very small, and systematic errors start to become more dominant. For example, the ideal CRB for SPLICE at $q=0.5$ with 320000 photons is, $\sqrt{2\pi e \sigma^2/N}= 0.007\sigma=3$ $\mu m$, which is comparable to the accuracy of our translation stages.

One can notice correlations between the RMSEs of $\delta$ and $q$ obtained from simulations at small $\delta$. The estimator for $\delta$ is very sensitive to the inaccuracies in the estimation of $q$. The RMSE of $q$ from the simulations of DI closely follow the CRB and saturate at a constant value. When the RMSE of $q$ becomes large due to systematic errors or $\tilde q$ becoming a biased estimator, RMSE of $\delta$ becomes worse than the CRB, which can be also seen in the Monte Carlo simulations for SPLICE.

The experimental RMSE for $q$ performs better than both the CRB for DI and the simulation results for all values of $q$. Surprisingly, the experimental RMSE in relative intensity is much lower than the QCRB for small separations. This is a parasitic effect of different cross-talks caused by the slight difference in the incidence angle of the beams. The background terms of $P_{\phi_\pm}$, $\epsilon_0+\epsilon_1\times q$, become the dominant term at small separations. The least-squares method exploits this information from $q$-dependent background and estimates $q$ better than the ideal case, it also affects the estimation of $\delta$ in a similar way. The values of $\epsilon_0$ and $\epsilon_1$ are $10^{-3}$ and $-2.1 \times 10^{-4}$ for $P_{\phi_-}$, and $2.4\times10^{-3}$ and $-10^{-3}$ for $P_{\phi_+}$, respectively.  Removing $\epsilon_1$ from the model worsens the estimates. Hence, it has to be included in the model. 
Calculating the FI for the background term of $P_{\phi_+}$ whose effect is the dominant one, yields a bound $\Delta q> \sqrt{F_q^{-1}}=\sqrt{\epsilon_0}/(\sqrt{N}\epsilon_1)= 0.11$, where $N=160{,}000$ for each projector. The RMSE for separation converges to this bound for small $\delta$. The RMSEs for larger separations $\delta /\sigma>0.2$ go much below $0.11$, which proves that the accuracy at this regime is due to SPLICE, not the $q$-dependent background.

\section{Conclusion}\label{sec:conc}
It is extremely difficult to accurately measure the separation or relative intensity of spatially unresolved sources.
It has long been established that SPLICE can improve the accuracy of separation measurement when the two sources are known to have equal intensities. Here we show that even when both the separation and the intensity imbalance are unknown, SPLICE can improve the accuracy of both parameters. Using only a fiber collimator and a phase shifter, SPLICE achieves an root mean squared error (RMSE) within approximately $50\%$ of the quantum limit. We demonstrate up to a sixfold reduction in RMSE for both parameters relative to intensity-based direct imaging (DI).

The maximum achievable advantage over DI is inversely proportional to the separation and to the square root of the intensity ratio. While the largest intensity ratio considered here was a factor of 9, for an intensity ratio of 100, one could expect an RMSE reduction factor of up to 60; and for an intensity ratio of $10^8$, the RMSE reduction would be around $6 \times 10^4$ for an ideal setup.
The observed advantage is limited in part by the finite crosstalk ($\epsilon\approx 10^{-3}$) of the mode projection; the maximum advantage is bounded by approximately $\sqrt{\rm 3/(2e\pi^2  \epsilon)} \approx 7.5$ in our experiment.  With better mode-sorting technology, this advantage would be higher; for instance, achieving an advantage of $40$ would require an extinction ratio of $10^6$ instead of $10^3$.

The diffraction limit remains a fundamental constraint on the resolution of conventional imaging systems. Increasing aperture size to mitigate diffraction can be prohibitively expensive. Phase-sensitive measurement techniques such as SPLICE offer a practical alternative and may substantially enhance imaging performance, particularly in applications such as exoplanet detection, where sources can exhibit extreme intensity imbalances.

\begin{backmatter}
\bmsection{Funding}
This work was supported by the Natural Sciences and Engineering Research Council (NSERC) of Canada,  CIFAR, QuEnSi, and the Fetzer Franklin Fund, a donor-advised fund of the Silicon Valley Community Foundation.

\bmsection{Acknowledgment}
We thank Edwin Tham, Hugo Ferretti, Jerry Tang, and Pria Dobney for their support and Mankei Tsang for the helpful conversations.  

\bmsection{Disclosures}
The authors declare no conflicts of interest.

\bmsection{Data Availability} Data underlying the results presented in this paper are not publicly available at this time but may
be obtained from the corresponding author upon reasonable request.

\end{backmatter}

%%%%%%%%%%%%%%%%%%%%%%% References %%%%%%%%%%%%%%%%%%%%%%%%%

%%%%%%%%%% If using BibTeX:
\bibliography{refs.bib}

@phdthesis{ferretti2022quantum,
  title={Quantum parameter estimation in the laboratory},
  author={Ferretti, Hugo},
  year={2022},
  school={University of Toronto (Canada)}
}

@article{rayleigh1879xxxi,
  title={XXXI. Investigations in optics, with special reference to the spectroscope},
  author={Rayleigh},
  journal={The London, Edinburgh, and Dublin Philosophical Magazine and Journal of Science},
  volume={8},
  number={49},
  pages={261--274},
  year={1879},
  publisher={Taylor \& Francis}
}

@article{abbe1873beitrage,
  title={Beitr{\"a}ge zur Theorie des Mikroskops und der mikroskopischen Wahrnehmung},
  author={Abbe, Ernst},
  journal={Archiv f{\"u}r mikroskopische Anatomie},
  volume={9},
  number={1},
  pages={413--468},
  year={1873},
  publisher={Springer}
}

@article{abbe1882relation,
  title={The Relation of Aperture and Power in the Microscope (continued).},
  author={Abbe},
  journal={Journal of the Royal Microscopical Society},
  volume={2},
  number={4},
  pages={460--473},
  year={1882},
  publisher={Blackwell Publishing Ltd Oxford, UK}
}

@inproceedings{schmidt1997telescope,
  title={Telescope costs and cost reduction},
  author={Schmidt-Kaler, Theodor and Rucks, Peter},
  booktitle={Optical Telescopes of Today and Tomorrow},
  volume={2871},
  pages={635--640},
  year={1997},
  organization={SPIE}
}

@article{hell1994breaking,
  title={Breaking the diffraction resolution limit by stimulated emission: stimulated-emission-depletion fluorescence microscopy},
  author={Hell, Stefan W and Wichmann, Jan},
  journal={Optics letters},
  volume={19},
  number={11},
  pages={780--782},
  year={1994},
  publisher={Optica Publishing Group}
}

@article{betzig2006imaging,
  title={Imaging intracellular fluorescent proteins at nanometer resolution},
  author={Betzig, Eric and Patterson, George H and Sougrat, Rachid and Lindwasser, O Wolf and Olenych, Scott and Bonifacino, Juan S and Davidson, Michael W and Lippincott-Schwartz, Jennifer and Hess, Harald F},
  journal={science},
  volume={313},
  number={5793},
  pages={1642--1645},
  year={2006},
  publisher={American Association for the Advancement of Science}
}

@article{hess2006ultra,
  title={Ultra-high resolution imaging by fluorescence photoactivation localization microscopy},
  author={Hess, Samuel T and Girirajan, Thanu PK and Mason, Michael D},
  journal={Biophysical journal},
  volume={91},
  number={11},
  pages={4258--4272},
  year={2006},
  publisher={Elsevier}
}

@article{hemmer2012universal,
  title={The universal scaling laws that determine the achievable resolution in different schemes for super-resolution imaging},
  author={Hemmer, Philip R and Zapata, Todd},
  journal={Journal of Optics},
  volume={14},
  number={8},
  pages={083002},
  year={2012},
  publisher={IOP Publishing}
}

@article{shin2011quantum,
  title={Quantum spatial superresolution by optical centroid measurements},
  author={Shin, Heedeuk and Chan, Kam Wai Clifford and Chang, Hye Jeong and Boyd, Robert W},
  journal={Physical review letters},
  volume={107},
  number={8},
  pages={083603},
  year={2011},
  publisher={APS}
}

@article{Tamburini2006vortice,
  title = {Overcoming the Rayleigh Criterion Limit with Optical Vortices},
  author = {Tamburini, F. and Anzolin, G. and Umbriaco, G. and Bianchini, A. and Barbieri, C.},
  journal = {Phys. Rev. Lett.},
  volume = {97},
  issue = {16},
  pages = {163903},
  numpages = {4},
  year = {2006},
  month = {Oct},
  publisher = {American Physical Society},
  doi = {10.1103/PhysRevLett.97.163903},
  url = {https://link.aps.org/doi/10.1103/PhysRevLett.97.163903}
}

@article{Tsang2009centroid,
  title = {Quantum Imaging beyond the Diffraction Limit by Optical Centroid Measurements},
  author = {Tsang, Mankei},
  journal = {Phys. Rev. Lett.},
  volume = {102},
  issue = {25},
  pages = {253601},
  numpages = {4},
  year = {2009},
  month = {Jun},
  publisher = {American Physical Society},
  doi = {10.1103/PhysRevLett.102.253601},
  url = {https://link.aps.org/doi/10.1103/PhysRevLett.102.253601}
}

@article{rozema2014scalable,
  title={Scalable spatial superresolution using entangled photons},
  author={Rozema, Lee A and Bateman, James D and Mahler, Dylan H and Okamoto, Ryo and Feizpour, Amir and Hayat, Alex and Steinberg, Aephraim M},
  journal={Physical review letters},
  volume={112},
  number={22},
  pages={223602},
  year={2014},
  publisher={APS}
}

@article{Giovannetti2009sub,
  title = {Sub-Rayleigh-diffraction-bound quantum imaging},
  author = {Giovannetti, Vittorio and Lloyd, Seth and Maccone, Lorenzo and Shapiro, Jeffrey H.},
  journal = {Phys. Rev. A},
  volume = {79},
  issue = {1},
  pages = {013827},
  numpages = {4},
  year = {2009},
  month = {Jan},
  publisher = {American Physical Society},
  doi = {10.1103/PhysRevA.79.013827},
  url = {https://link.aps.org/doi/10.1103/PhysRevA.79.013827}
}

@article{schwartz2013superresolution,
  title={Superresolution microscopy with quantum emitters},
  author={Schwartz, Osip and Levitt, Jonathan M and Tenne, Ron and Itzhakov, Stella and Deutsch, Zvicka and Oron, Dan},
  journal={Nano letters},
  volume={13},
  number={12},
  pages={5832--5836},
  year={2013},
  publisher={ACS Publications}
}

@article{tsang2016quantum,
  title={Quantum theory of superresolution for two incoherent optical point sources},
  author={Tsang, Mankei and Nair, Ranjith and Lu, Xiao-Ming},
  journal={Physical Review X},
  volume={6},
  number={3},
  pages={031033},
  year={2016},
  publisher={APS}
}

@article{lupo2016ultimate,
  title={Ultimate precision bound of quantum and subwavelength imaging},
  author={Lupo, Cosmo and Pirandola, Stefano},
  journal={Physical review letters},
  volume={117},
  number={19},
  pages={190802},
  year={2016},
  publisher={APS}
}

@article{rehacek2017optimal,
  title={Optimal measurements for resolution beyond the Rayleigh limit},
  author={Rehacek, Jaroslav and Pa{\'u}r, Martin and Stoklasa, Bohumil and Hradil, Zdenek and Sanchez-Soto, Luis L},
  journal={Optics letters},
  volume={42},
  number={2},
  pages={231--234},
  year={2017},
  publisher={Optica Publishing Group}
}

@article{gefen2019overcoming,
  title={Overcoming resolution limits with quantum sensing},
  author={Gefen, Tuvia and Rotem, Amit and Retzker, Alex},
  journal={Nature communications},
  volume={10},
  number={1},
  pages={4992},
  year={2019},
  publisher={Nature Publishing Group UK London}
}

@article{tang2016fault,
  title={Fault-tolerant and finite-error localization for point emitters within the diffraction limit},
  author={Tang, Zong Sheng and Durak, Kadir and Ling, Alexander},
  journal={Optics express},
  volume={24},
  number={19},
  pages={22004--22012},
  year={2016},
  publisher={Optical Society of America}
}

@article{paur2016achieving,
  title={Achieving the ultimate optical resolution},
  author={Pa{\'u}r, Martin and Stoklasa, Bohumil and Hradil, Zdenek and S{\'a}nchez-Soto, Luis L and Rehacek, Jaroslav},
  journal={Optica},
  volume={3},
  number={10},
  pages={1144--1147},
  year={2016},
  publisher={Optica Publishing Group}
}

@article{tham2017beating,
  title={Beating Rayleigh’s curse by imaging using phase information},
  author={Tham, Weng-Kian and Ferretti, Hugo and Steinberg, Aephraim M},
  journal={Physical review letters},
  volume={118},
  number={7},
  pages={070801},
  year={2017},
  publisher={APS}
}

@article{donohue2018quantum,
  title={Quantum-limited time-frequency estimation through mode-selective photon measurement},
  author={Donohue, John M and Ansari, Vahid and {\v{R}}eh{\'a}{\v{c}}ek, Jaroslav and Hradil, Zdenek and Stoklasa, Bohumil and Pa{\'u}r, Martin and S{\'a}nchez-Soto, Luis L and Silberhorn, Christine},
  journal={Physical review letters},
  volume={121},
  number={9},
  pages={090501},
  year={2018},
  publisher={APS}
}

@article{howell2023super,
  title={Super interferometric range resolution},
  author={Howell, John C and Jordan, Andrew N and {\v{S}}oda, Barbara and Kempf, Achim},
  journal={Physical Review Letters},
  volume={131},
  number={5},
  pages={053803},
  year={2023},
  publisher={APS}
}

@article{frank2023passive,
  title={Passive superresolution imaging of incoherent objects},
  author={Frank, Jernej and Duplinskiy, Alexander and Bearne, Kaden and Lvovsky, AI},
  journal={Optica},
  volume={10},
  number={9},
  pages={1147--1152},
  year={2023},
  publisher={Optica Publishing Group}
}

@article{rouviere2024ultra,
  title={Ultra-sensitive separation estimation of optical sources},
  author={Rouvi{\`e}re, Cl{\'e}mentine and Barral, David and Grateau, Antonin and Karuseichyk, Ilya and Sorelli, Giacomo and Walschaers, Mattia and Treps, Nicolas},
  journal={Optica},
  volume={11},
  number={2},
  pages={166--170},
  year={2024},
  publisher={Optica Publishing Group}
}

@article{zhou2019quantum,
  title={Quantum-limited estimation of the axial separation of two incoherent point sources},
  author={Zhou, Yiyu and Yang, Jing and Hassett, Jeremy D and Rafsanjani, Seyed Mohammad Hashemi and Mirhosseini, Mohammad and Vamivakas, A Nick and Jordan, Andrew N and Shi, Zhimin and Boyd, Robert W},
  journal={Optica},
  volume={6},
  number={5},
  pages={534--541},
  year={2019},
  publisher={Optica Publishing Group}
}

@article{yang2016far,
  title={Far-field linear optical superresolution via heterodyne detection in a higher-order local oscillator mode},
  author={Yang, Fan and Tashchilina, Arina and Moiseev, Eugene S and Simon, Christoph and Lvovsky, Alexander I},
  journal={Optica},
  volume={3},
  number={10},
  pages={1148--1152},
  year={2016},
  publisher={Optica Publishing Group}
}

@article{tsang2017subdiffraction,
  title={Subdiffraction incoherent optical imaging via spatial-mode demultiplexing},
  author={Tsang, Mankei},
  journal={New Journal of Physics},
  volume={19},
  number={2},
  pages={023054},
  year={2017},
  publisher={IOP Publishing}
}

@book{traub2010direct,
  title={Direct imaging of exoplanets},
  author={Traub, Wesley A and Oppenheimer, Ben R},
  year={2010},
  publisher={University of Arizona Press, Tucson}
}

@article{huang2021quantum,
  title={Quantum hypothesis testing for exoplanet detection},
  author={Huang, Zixin and Lupo, Cosmo},
  journal={Physical Review Letters},
  volume={127},
  number={13},
  pages={130502},
  year={2021},
  publisher={APS}
}

@article{deshler2024achieving,
  title={Achieving Quantum Limits of Exoplanet Detection and Localization},
  author={Deshler, Nico and Haffert, Sebastiaan and Ashok, Amit},
  journal={arXiv preprint arXiv:2403.17988},
  year={2024}
}

@article{radhakrishnan2025towards,
  title={Towards real-time contrast control for direct exoplanet imaging with adaptive optics},
  author={Radhakrishnan, Vikram M and Keller, Christoph U and Doelman, Niek J},
  journal={Optics Communications},
  volume={574},
  pages={131046},
  year={2025},
  publisher={Elsevier}
}

@article{helstrom1967minimum,
  title={Minimum mean-squared error of estimates in quantum statistics},
  author={Helstrom, Carl W},
  journal={Physics letters A},
  volume={25},
  number={2},
  pages={101--102},
  year={1967},
  publisher={Elsevier}
}

@article{Soto2017multiparam,
  title = {Multiparameter quantum metrology of incoherent point sources: Towards realistic superresolution},
  author = {\ifmmode \check{R}\else \v{R}\fi{}eha\ifmmode \check{c}\else \v{c}\fi{}ek, J. and Hradil, Z. and Stoklasa, B. and Pa\'ur, M. and Grover, J. and Krzic, A. and S\'anchez-Soto, L. L.},
  journal = {Phys. Rev. A},
  volume = {96},
  issue = {6},
  pages = {062107},
  numpages = {7},
  year = {2017},
  month = {Dec},
  publisher = {American Physical Society},
  doi = {10.1103/PhysRevA.96.062107},
  url = {https://link.aps.org/doi/10.1103/PhysRevA.96.062107}
}

@article{bonsma2019realistic,
  title={Realistic sub-Rayleigh imaging with phase-sensitive measurements},
  author={Bonsma-Fisher, Kent AG and Tham, Weng-Kian and Ferretti, Hugo and Steinberg, Aephraim M},
  journal={New Journal of Physics},
  volume={21},
  number={9},
  pages={093010},
  year={2019},
  publisher={IOP Publishing}
}

@article{PhysRevA.100.032104,
  title = {Optimal measurements for quantum multiparameter estimation with general states},
  author = {Yang, Jing and Pang, Shengshi and Zhou, Yiyu and Jordan, Andrew N.},
  journal = {Phys. Rev. A},
  volume = {100},
  issue = {3},
  pages = {032104},
  numpages = {14},
  year = {2019},
  month = {Sep},
  publisher = {American Physical Society},
  doi = {10.1103/PhysRevA.100.032104},
  url = {https://link.aps.org/doi/10.1103/PhysRevA.100.032104}
}

@article{prasad2020quantum,
  title={Quantum limited super-resolution of an unequal-brightness source pair in three dimensions},
  author={Prasad, Sudhakar},
  journal={Physica Scripta},
  volume={95},
  number={5},
  pages={054004},
  year={2020},
  publisher={IOP Publishing}
}

@article{xin2021optical,
  title={Optical super-resolution for two unequally bright point sources based on the fractional Hilbert transform},
  author={Xin, Jun and Li, Yanan and Lu, Xiao-Ming},
  journal={Physical Review A},
  volume={103},
  number={5},
  pages={052604},
  year={2021},
  publisher={APS}
}

@techreport{katamadze2023breaking,
  title={Breaking Rayleigh’s curse for two unbalanced single-photon emitters using BLESS technique},
  author={Katamadze, Konstantin and Bantysh, Boris and Chernyavskiy, Andrey and Bogdanov, Yurii and Kulik, Sergei},
  year={2023},
  institution={Optica Open}
}

@article{ansari2021achieving,
  title={Achieving the ultimate quantum timing resolution},
  author={Ansari, Vahid and Brecht, Benjamin and Gil-Lopez, Jano and Donohue, John M and {\v{R}}eh{\'a}{\v{c}}ek, Jaroslav and Hradil, Zden{\v{e}}k and S{\'a}nchez-Soto, Luis L and Silberhorn, Christine},
  journal={PRX Quantum},
  volume={2},
  number={1},
  pages={010301},
  year={2021},
  publisher={APS}
}

@article{santamaria2023spatial,
  title={Spatial-mode demultiplexing for enhanced intensity and distance measurement},
  author={Santamaria, Luigi and Pallotti, Deborah and de Cumis, Mario Siciliani and Dequal, Daniele and Lupo, Cosmo},
  journal={Optics Express},
  volume={31},
  number={21},
  pages={33930--33944},
  year={2023},
  publisher={Optica Publishing Group}
}

@article{santamaria2024single,
  title={Single-photon sub-Rayleigh precision measurements of a pair of incoherent sources of unequal intensity},
  author={Santamaria, Luigi and Sgobba, Fabrizio and Lupo, Cosmo},
  journal={Optica Quantum},
  volume={2},
  number={1},
  pages={46--56},
  year={2024},
  publisher={Optica Publishing Group}
}

@article{wallis2025spatial,
  title={Spatial mode demultiplexing for super-resolved source parameter estimation},
  author={Wallis, John S and Gozzard, David R and Frost, Alex M and Collier, Joshua J and Maron, Nicolas and Dix-Matthews, Benjamin P},
  journal={Optics Express},
  volume={33},
  number={16},
  pages={34651--34662},
  year={2025},
  publisher={Optica Publishing Group}
}

@article{grateau2026multiparameter,
  title={Multiparameter estimation for the superresolution of two incoherent sources},
  author={Grateau, Antonin and Boeschoten, Alexander and Favin-L{\'e}v{\^e}que, Tanguy and Herrera, Isael and Treps, Nicolas},
  journal={arXiv preprint arXiv:2601.14876},
  year={2026}
}

@article{rao1945information,
  title={Information and the accuracy attainable in the estimation of statistical parameters},
  author={Rao, C Radhakrishna and others},
  journal={Bull. Calcutta Math. Soc},
  volume={37},
  number={3},
  pages={81--91},
  year={1945}
}

@article{ly2017tutorial,
  title={A tutorial on Fisher information},
  author={Ly, Alexander and Marsman, Maarten and Verhagen, Josine and Grasman, Raoul PPP and Wagenmakers, Eric-Jan},
  journal={Journal of Mathematical Psychology},
  volume={80},
  pages={40--55},
  year={2017},
  publisher={Elsevier}
}

@article{helstrom1969quantum,
  title={Quantum detection and estimation theory},
  author={Helstrom, Carl W},
  journal={Journal of Statistical Physics},
  volume={1},
  pages={231--252},
  year={1969},
  publisher={Springer}
}

@article{liu2020quantum,
  title={Quantum Fisher information matrix and multiparameter estimation},
  author={Liu, Jing and Yuan, Haidong and Lu, Xiao-Ming and Wang, Xiaoguang},
  journal={Journal of Physics A: Mathematical and Theoretical},
  volume={53},
  number={2},
  pages={023001},
  year={2019},
  publisher={IOP Publishing}
}

@book{tham2020quantum,
  title={Quantum Homomorphic Encryption: Implementation and Applications},
  author={Tham, Weng-Kian Edwin},
  year={2020},
  publisher={University of Toronto (Canada)}
}

@article{giovannetti2004quantum,
  title={Quantum-enhanced measurements: beating the standard quantum limit},
  author={Giovannetti, Vittorio and Lloyd, Seth and Maccone, Lorenzo},
  journal={Science},
  volume={306},
  number={5700},
  pages={1330--1336},
  year={2004},
  publisher={American Association for the Advancement of Science}
}

\end{document}